\documentclass[twocolumn,prb,aps,epsf,showpacs,superscriptaddress]{revtex4-1}

\usepackage{graphicx}
\usepackage{dcolumn}
\usepackage{amsmath}
\usepackage{bm}
\usepackage{lipsum}

\newcommand{\D}{\mathrm{d}}
\newcommand{\I}{\mathrm{i}}
\newcommand{\lsco}{La$_{2-x}$Sr$_x$CuO$_4$}
\newcommand{\tbcod}{Tl$_2$Ba$_2$CuO$_{6+\delta}$}

\begin{document}

\title{Disorder and  superfluid density in overdoped cuprate superconductors}

\author{N.~R.~Lee-Hone}
\affiliation{Department of Physics, Simon Fraser University, Burnaby, BC, V5A~1S6, Canada}
\author{J.~S.~Dodge}
\author{D.~M.~Broun}
\affiliation{Department of Physics, Simon Fraser University, Burnaby, BC, V5A~1S6, Canada}
\affiliation{Canadian Institute for Advanced Research, Toronto, ON, MG5 1Z8, Canada}

\begin{abstract}
We calculate superfluid density for a dirty $d$-wave superconductor.  The effects of impurity scattering are treated within the self-consistent $t$-matrix approximation, in weak-coupling BCS theory.  Working from a realistic tight-binding parameterization of the Fermi surface, we find a superfluid density that is both correlated with $T_c$ and linear in temperature, in good correspondence with recent experiments on overdoped \lsco.
\end{abstract}
 
 
\maketitle{} 
  
\section{Introduction}

The superfluid density, $\rho_s$, plays a special role in the physics of cuprate superconductors, as it determines the stiffness of the superconducting order parameter to fluctuations in its phase.\cite{Emery1995,Orenstein2000}
In most superconductors $\rho_s$ is large: phase fluctuations are heavily suppressed and the transition temperature $T_c$ is set primarily by the gap energy $2\Delta$ required to break a Cooper pair. Cuprates, on the other hand, have a relatively low carrier density, which limits  $\rho_s$ and leaves them  susceptible to phase fluctuations. As a result, both $\Delta$ and $\rho_s$ may influence $T_c$ in the cuprates, with phase fluctuations playing an increasingly dominant role as $\rho_s$ approaches zero. Experiments support this view:
in hole-doped cuprates in the underdoped regime, $T_c$ correlates closely with $\rho_s$,\cite{Uemura1989,Liang:2005p493,Hetel:2007p1005,Broun:2007p49,Sonier2016} whereas there is a large energy gap that extends well into the normal state.\cite{Timusk:1999p422,Sutherland:2003p753,Hufner2008}  In addition, a variety of superconducting fluctuation effects have been observed above $T_c$ in underdoped cuprates.\cite{Corson:1999p716,Xu:2000p609,Wang:2002p645,Wang:2005p2400,Wang:2006p185,Kitano:2006id,Li:2007p180,Grbic:2009kd,Rourke2011,Bilbro:2011jj,Grbic:2011hp,Behnia:2016fc} On the overdoped side the situation is different: $T_c$ appears to closely track the energy gap,\cite{Sutherland:2003p753,Hawthorn:2007hq,Hufner2008} as it would in a conventional superconductor; and, while the correlation between $T_c$ and $\rho_s$ remains,\cite{Uemura1993,Niedermayer1993,Locquet1996,Bernhard2001,Wang2007,Lemberger2011,Bozovic:2016ei} the causal relationship between these quantities is far less clear.  Complicating the chain of causality is another parameter --- disorder --- that directly influences $\rho_s$, $T_c$, and $\Delta$ in a $d$-wave superconductor.\cite{Puchkaryov1998,PROHAMMER:1991p557,Hirschfeld:1993cka,Hussey:2002p649,Alloul:2009dk,Kogan:2013kf} One of our main purposes in this paper is to explore the extent to which disorder is an important driver of the relationships between the other three quantities, with particular attention to the case of the overdoped cuprates.

Strong motivation comes from a recent experiment,\cite{Bozovic:2016ei} which provides exhaustive evidence that the superfluid stiffness and the superconducting transition temperature of overdoped \lsco\ approach zero in tandem as a function of doping.  The new study  reinforces the argument that the close correlation between $T_c$ and $\rho_s$ is a significant and intrinsic feature of the overdoped cuprates.  It also shows that the superfluid density retains an approximately linear temperature dependence over a wide temperature range, as expected for a $d$-wave superconductor in the clean limit.\cite{Lemberger2010,Bozovic:2016ei} Together, these observations present a puzzle:  on the one hand the superfluid stiffness is expected to correlate with $T_c$ in the \textit{dirty} limit, because the normal-state spectral weight is cut off by the gap,\cite{Mattis1958,Abrikosov1959,Puchkaryov1998,Lemberger2011} but on the other hand the observed temperature dependence of $\rho_s$ appears to exclude this possibility. 

To try to resolve this contradiction we have revisited the theoretical relationship between disorder, superfluid density and $T_c$ within  dirty $d$-wave BCS theory.\cite{PROHAMMER:1991p557,Hirschfeld:1993cka,Puchkaryov1998} In a dirty $d$-wave superconductor it is well known that strong-scattering (unitarity-limit) impurities rapidly induce a crossover from $T$-linear superfluid density to quadratic behaviour\cite{PROHAMMER:1991p557,Hirschfeld:1993cka,Kogan:2013kf}  below a crossover temperature $T^\ast$ that is proportional to the geometric mean of the normal-state impurity scattering rate and the superconducting energy gap.\cite{Hirschfeld:1993cka}
The corresponding loss of superfluid density is of order $T^\ast\!/T_c$.  Therefore, in this limit, any significant loss of superfluid density must be accompanied by a very visible crossover to quadratic behaviour in $\rho_s(T)$.  This result is so well known that the argument is frequently run in reverse, with the measured value of $T^\ast$ used to place an upper bound on the degree of superfluid suppression and $T_c$ suppression due to impurities.  In fact, we will show that the reverse argument breaks down for weak-scattering (Born-limit) disorder, and approximately linear-in-temperature superfluid density can coexist with substantial suppression of superfluid density and $T_c$.  We have carried out calculations of  superfluid density for realistic, doping-dependent Fermi surfaces based on tight-binding parameterizations of angle-resolved photoemission (ARPES) dispersions for \lsco.\cite{Yoshida:2006hw,Razzoli:2010hi} This turns out to be crucial to carrying out a detailed comparison with $\rho_s(T)$ data on \lsco.  In the calculations, the effects of disorder on the quasiparticle energies and lifetimes, and on the superconducting energy gap and $T_c$, are calculated using the self-consistent $t$-matrix approximation, within the weak-coupling limit of $d$-wave BCS theory.\cite{PROHAMMER:1991p557,Hirschfeld:1993cka}   We conclude that it is possible to obtain a superfluid density that is both correlated with $T_c$ and linear in temperature.

\section{Theory}

\label{theory}

\subsection{Dirty d-wave superconductivity}

The gap equation for a weak-coupling $d$-wave superconductor can be written in the imaginary-axis formalism as\cite{PROHAMMER:1991p557}
\begin{equation}
\Delta_\mathbf{k} =2 \pi T \sum_{\omega_n > 0}^{\omega_0}\!\left\langle V_{\mathbf{k},\mathbf{k^\prime}} \frac{\Delta_\mathbf{k^\prime}}{\sqrt{\tilde \omega_n^2 + \Delta_\mathbf{k^\prime}^2}}\right\rangle_{\!\!\mathrm{FS}},
\end{equation}
where $\Delta_\mathbf{k}$ is the gap parameter at wave-vector $\mathbf{k}$, \mbox{$\omega_n = 2 \pi T (n + \tfrac{1}{2})$} are the fermionic  Matsubara frequencies, $V_{\mathbf{k},\mathbf{k^\prime}}$ is the  pairing interaction, $\omega_0$ is a high frequency cut off, and $\langle ... \rangle_\mathrm{FS}$ denotes an average over the Fermi surface.  

In the self-consistent $t$-matrix approximation,\cite{PROHAMMER:1991p557,Hirschfeld:1993cka} point-like, nonmagnetic impurities renormalize the fermionic Matsubara frequencies according to  
\begin{equation}
\tilde \omega_n \equiv \tilde \omega(\omega_n) = \omega_n + \pi \Gamma \frac{\langle N_\mathbf{k}(\tilde \omega_n) \rangle_\mathrm{FS}}{c^2 + \langle N_\mathbf{k}(\tilde \omega_n) \rangle_\mathrm{FS}^2}\;.
\label{tmatrix}
\end{equation}
Here \mbox{$c$ is the cotangent} of the scattering phase shift, $\Gamma$ is a scattering parameter proportional to the concentration of impurities and
\begin{equation}
N_\mathbf{k}(\tilde \omega_n) = \frac{\tilde \omega_n}{\sqrt{\tilde \omega_n^2 + \Delta_\mathbf{k}^2}}\;.
\end{equation}
For a $d$-wave order parameter, which averages to zero over the Fermi surface, there is no explicit impurity renormalization of $\Delta_\mathbf{k}$, just an indirect reduction through the effect of the impurities on $\tilde \omega_n$.

For simplicity, we assume a separable pairing interaction based on a $d$-wave form factor
$\Omega_\mathbf{k}$ defined in the first Brillouin zone of the two-dimensional CuO$_2$ planes,
\begin{equation}
\Omega_\mathbf{k} \propto \big(\cos(k_x a) - \cos(k_y a) \big)\;,
\end{equation}
where $a$ is the lattice spacing and $\Omega_\mathbf{k}$ is normalized such that $\langle \Omega_\mathbf{k}^2\rangle_\mathrm{FS} = 1$.  The pairing interaction therefore takes the form \begin{equation}
V_{\mathbf{k},\mathbf{k^\prime}} = V_0 \Omega_\mathbf{k} \, \Omega_\mathbf{k^\prime}\;.
\end{equation}
We will see below that for weak-coupling BCS, where the cut-off frequency of the interaction, $\omega_0$, is much larger than the superconducting transition temperature, $T_c$, the combined effect of $V_0$ and $\omega_0$ is captured by the clean-limit transition temperature, $T_{c0}$, so that  $V_0$ and $\omega_0$ do not appear explicitly as parameters in the theory.  Introducing a temperature-dependent gap amplitude, $\psi(T)$, the gap equation becomes
\begin{equation}
\Delta_\mathbf{k} \equiv  \psi \Omega_\mathbf{k} = 2 \pi T \!\!\sum_{\omega_n > 0}^{\omega_0} \!\!\left\langle \!V_0\Omega_\mathbf{k}\Omega_\mathbf{k^\prime} \frac{\psi \Omega_\mathbf{k^\prime}}{(\tilde \omega_n^2 + \psi^2 \Omega_\mathbf{k^\prime}^2)^\frac{1}{2}}\!\right\rangle_{\!\!\!\mathrm{FS}}\!\!.
\end{equation}

Cancelling common factors, rearranging and reassigning \mbox{$\mathbf{k^\prime} \to \mathbf{k}$} we have
\begin{equation}
\frac{1}{V_0} = 2 \pi T  \sum_{\omega_n > 0}^{\omega_0} \left\langle \frac{\Omega_\mathbf{k}^2}{(\tilde \omega_n^2 + \psi^2\Omega_\mathbf{k}^2)^\frac{1}{2}} \right\rangle_{\!\mathrm{FS}}\,.
\end{equation}
In the absence of disorder the quasiparticle energies are unrenormalized ($\tilde \omega_n = \omega_n$) and the gap vanishes  \mbox{($\psi \to 0$)} at the clean-limit transition temperature, $T_{c0}$.  Using $\langle \Omega_\mathbf{k}^2\rangle_\mathrm{FS} = 1$, we have at this temperature
\begin{equation}
\frac{1}{V_0} = 2 \pi T_{c0}  \sum_{\omega_n > 0}^{\omega_0} \frac{1}{\omega_n(T_{c0})} \approx \ln\left(\frac{2 \omega_0}{1.76\, T_{c0}}\right)\;,
\label{BCS_Tc}
\end{equation}
where the approximation is valid when $\omega_0 \gg T_{c0}$.  This rearranges to give the familiar weak-coupling BCS result
\begin{equation}
T_{c0} = 1.14\, \omega_0 \exp(-1/V_0)\;.
\end{equation}
The logarithmic temperature dependence in Eq.~\ref{BCS_Tc} can be used to obtain an expression for the coupling constant $V_0$ that applies at any arbitrary temperature $T$:\cite{Eilenberger:1968bb,Kogan:2009ew}
\begin{equation}
\frac{1}{V_0} = 2 \pi T \sum_{\omega_n > 0}^{\omega_0} \frac{1}{\omega_n(T)}  + \ln\left(\frac{T}{T_{c0}}\right)\;.
\end{equation}
This allows $V_0$ to be eliminated from the gap equation, which then takes the form
\begin{equation}
\ln\left(\!\frac{T_{c0}}{T}\!\right) = 2 \pi T \!\sum_{\omega_n > 0}^\infty \!\left(\frac{1}{\omega_n} \!-\! \left\langle\!\frac{\Omega_\mathbf{k}^2}{(\tilde \omega_n^2 + \psi^2\Omega_\mathbf{k}^2)^\frac{1}{2}}\! \right\rangle_{\!\!\!\mathrm{FS}} \right).
\label{gap_equation}
\end{equation}
Rapid convergence  lets the Matsubara sum to be taken to infinity, eliminating  explicit dependence on $\omega_0$.  For a given choice of Fermi surface and impurity parameters, Eqs.~\ref{tmatrix} and \ref{gap_equation} are solved self consistently to obtain $\tilde \omega_n(T)$ and $\psi(T)$. 

In the presence of disorder the energy gap closes at a reduced transition temperature, $T_c$.  For $T \ge T_c$, $N_\mathbf{k}(\tilde \omega) \to 1$ and the $t$-matrix equation describing the impurity scattering, Eq.~\ref{tmatrix}, simplifies to
\begin{equation}
 \tilde \omega(\omega_n) = \omega_n +  \frac{\pi \Gamma}{1 + c^2} \equiv \omega_n + \Gamma_N\,.
\label{norm_state_gamma}
\end{equation}
The imaginary part of the self energy in this limit is denoted $\Gamma_N$, the normal-state scattering rate due to impurities. Equation~\ref{gap_equation} can be solved with $\psi \to 0$ and $\tilde\omega_n \to \omega_n + \Gamma_N$ to determine $T_c$:
\begin{align}
\ln\left(\!\frac{T_{c0}}{T_{c}}\!\right) & = 2 \pi T_c \!\sum_{\omega_n > 0}^\infty \!\left(\frac{1}{\omega_n} - \frac{1}{\omega_n+\Gamma_N} \right)\\
& = \sum_{\omega_n > 0}^\infty \!\left(\frac{1}{n + \tfrac{1}{2}} - \frac{1}{n + \tfrac{1}{2}+\frac{\Gamma_N}{2 \pi T_c}} \right)\\
& = \psi_0\!\left(\tfrac{1}{2}+ \frac{\Gamma_N}{2 \pi T_c}\right) - \psi_0\!\left(\tfrac{1}{2} \right)\;,
\label{Tc_equation}
\end{align}
where $\psi_0(x)$ is the digamma function.

\subsection{Superfluid density}

The zero-temperature, zero-disorder penetration depth, $\lambda_{00}$, is closely related to the bare plasma frequency, $\omega_p$.   The corresponding superfluid density is\cite{CHANDRASEKHAR:1993p329}
\begin{align}
\rho_{s00} & \equiv \frac{1}{\lambda_{00}^2} = \mu_0 \epsilon_0 \omega_p^2\\
&  = 2 \mu_0 e^2 \int_{-\frac{\pi}{d}}^{+\frac{\pi}{d}}\frac{\D k_z}{2 \pi}\int \frac{\D^2 \mathbf{k}}{(2 \pi)^2} \delta(\epsilon_F - \epsilon_\mathbf{k}) {v}_{\mathbf{k},x}^2\,.
\label{plasmaFreq1}
\end{align}
Here we specialize to a quasi-2D material with layer spacing $d$ and in-plane energy dispersion $\epsilon_\mathbf{k}$. $\epsilon_F$ is the Fermi energy, $\mathbf{k} = (k_x,k_y)$ is the in-plane momentum and $\mathbf{v}_\mathbf{k} = \frac{1}{\hbar}\big(\frac{\partial}{\partial k_x},\frac{\partial}{\partial k_y}\big)\epsilon_\mathbf{k}$ is the in-plane velocity.  We change Eq.~\ref{plasmaFreq1} to a Fermi surface integral by transforming coordinates from $(k_x,k_y)$ to $(\epsilon,\phi)$, where $\phi$ is the angle in the plane, measured about  $\left(\frac{\pi}{a},\frac{\pi}{a}\right)$ at low hole dopings and $(0,0)$ at higher dopings.  The Jacobian of the transformation is
\begin{equation}
J(\phi) = \frac{\partial(k_x,k_y)}{\partial(\epsilon,\phi)} = \frac{|k|^2}{\hbar\, \mathbf{k}\!\cdot\!\mathbf{v}_\mathbf{k}}\;.
\end{equation}
When the energy and $k_z$ integrations are carried out we obtain
\begin{equation}
\frac{1}{\lambda_{00}^2} = \frac{\mu_0 e^2}{2 \pi^2 \hbar d} \int_0^{2\pi} \frac{|k_F|^2}{\mathbf{k}_F\!\cdot\!\mathbf{v}_F} {v}_{F,x}^2 \D \phi\;,
\end{equation}
where the Fermi wavevector $\mathbf{k}_F$ and Fermi velocity $\mathbf{v}_F$ are functions of $\phi$.  The Fermi surface average used in the previous section, $\langle ... \rangle_\mathrm{FS}$, must include the same Jacobian factor:
\begin{equation}
\big\langle A(\phi) \big\rangle_\mathrm{FS} \equiv \int_0^{2\pi} \!\!\!J(\phi) A(\phi) \D \phi\Big/\!\!\int_0^{2\pi} \!\!\!J(\phi) \D \phi\;.
\end{equation}
For calculation of plasma frequency and superfluid density we define a second Fermi surface average, $\langle\!\langle ... \rangle\!\rangle_\mathrm{FS}$, that contains the additional factor of ${v}_{F,x}^2$:
\begin{equation}
\big\langle\!\big\langle A(\phi) \big\rangle\!\big\rangle_\mathrm{FS} \equiv \int_0^{2\pi} \!\!\!\!\!J(\phi) A(\phi) {v}_{F,x}^2 \D \phi\Big/\!\!\int_0^{2\pi} \!\!\!\!\!J(\phi)  {v}_{F,x}^2\D \phi\,.
\end{equation}
For Fermi surfaces that are close to circular these distinctions are usually not important.  However, for the overdoped cuprates, the details of the Fermi-surface averages turn out to be crucial to understanding the temperature dependence of superfluid density measured in experiments, and so are given here in full.  
\begin{figure}[t]
\centering
\includegraphics[width = 0.9 \columnwidth]{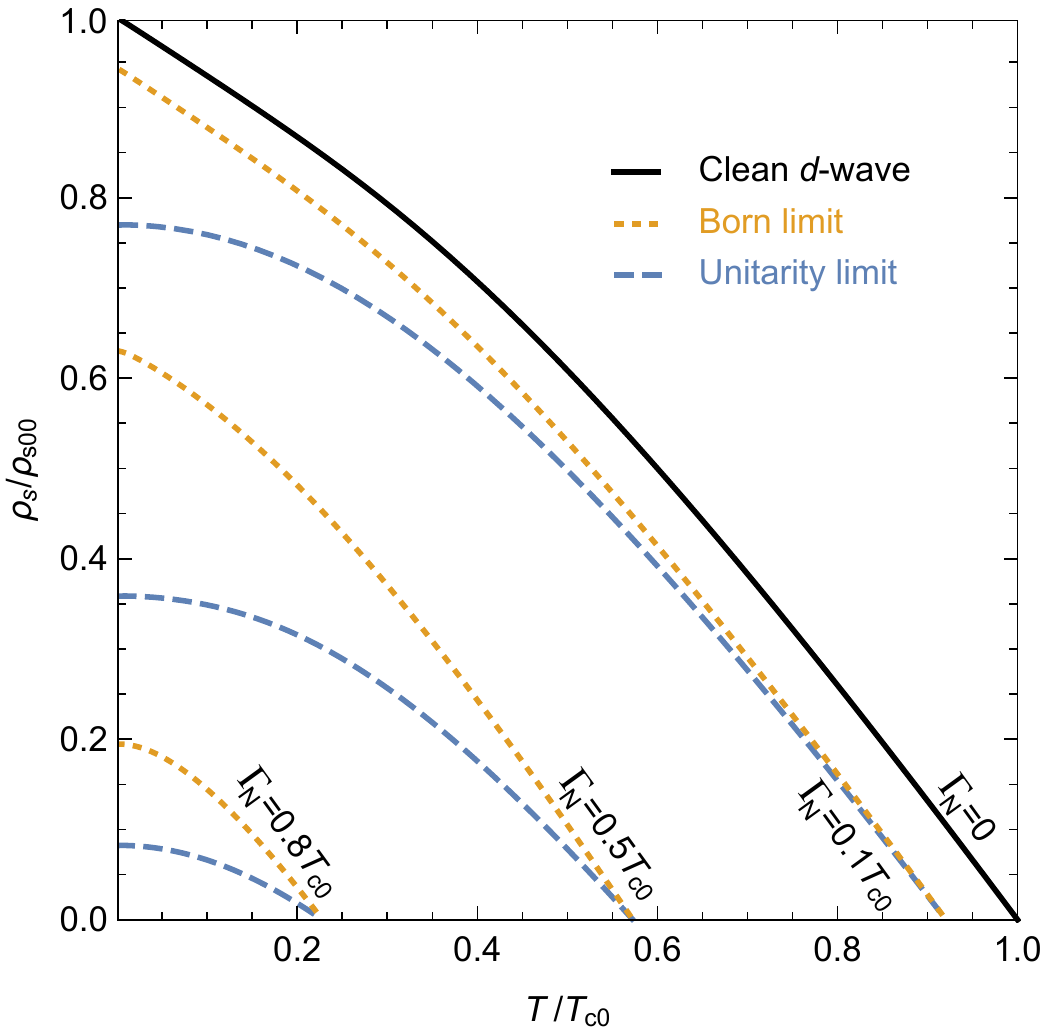}
\caption{Normalized superfluid density, $\rho_s/\rho_{s00}$, for a $d$-wave superconductor with a circular Fermi surface.  The degree of scattering is characterized by the normal-state scattering rate, $\Gamma_N$, in units of the clean limit transition temperature, $T_{c0}$, for scatterers acting in the Born limit ($c \gg 1$) and the unitarity limit  ($c = 0$).  The temperature dependence of the gap, $\Delta(T)$, has been calculated self-consistently for each set of impurity parameters, assuming a separable $d$-wave pairing interaction.}
\label{fig:CircularFS}
\end{figure} 
\begin{figure*}[t]
\centering
\includegraphics[width = \textwidth]{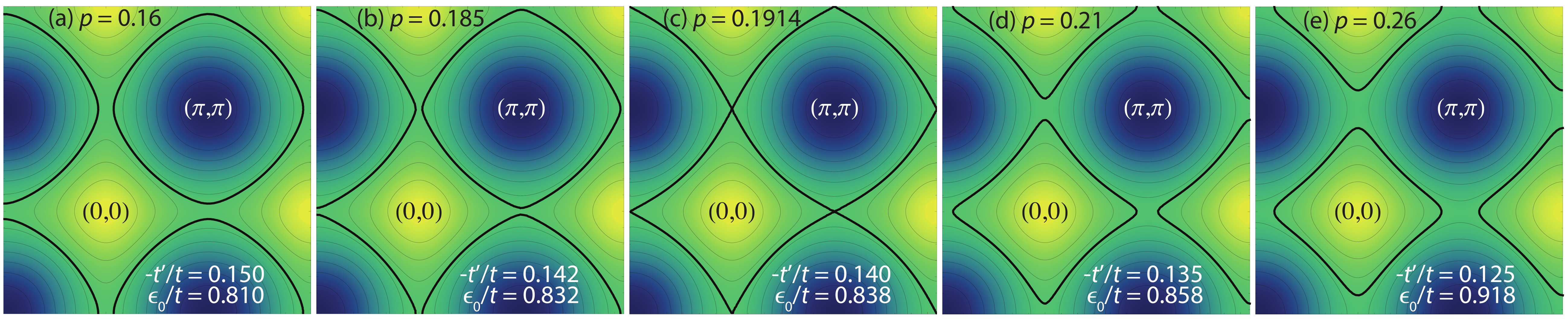}
\caption{(color online) Constant energy contours in momentum space for optimally to overdoped \lsco, at selected nominal hole-dopings $p$, based on a tight-binding parameterization of ARPES spectra.\cite{Yoshida:2006hw} Momentum is measured in units of inverse lattice spacing, $1/a$. Fermi surfaces are depicted by solid black lines.  Doping-dependent tight-binding parameters $t^\prime$ and $\epsilon_0$ are indicated on the plots, in units of nearest-neighbour hopping integral $t$.} 
\label{fig:FermiSurfaces}
\end{figure*} 

The finite temperature superfluid density, in the presence of disorder, is most efficiently calculated using a Matsubara sum.\cite{NAM:1967p640}  Normalized to $\rho_{s00}$ it is given by 
\begin{equation}
\frac{\rho_s(T)}{\rho_{s00}} = 2 \pi T \sum_{\omega_n > 0}^\infty \left\langle\!\!\! \left\langle \frac{\Delta_\mathbf{k}^2}{(\tilde \omega_n^2 + \Delta_\mathbf{k}^2 )^\frac{3}{2}} \right\rangle\!\!\!\right\rangle_{\!\!\mathrm{FS}}\,,
\label{superfluiddensity}
\end{equation}
where the effects of disorder are built in via the renormalized Matsubara frequencies and gap.

\subsection{Impurity contribution to normal-state resistivity}

\label{resistivity}

We use the normal-state impurity scattering rate $\Gamma_N$ to parameterize the amount of scattering in the theory.  Motivated by the known types of elastic-scattering disorder in \lsco\ and other cuprates, we allow for two types of defects acting in combination: weak-limit scatterers, parameterized by $\Gamma_{N,\mathrm{Born}}$, to capture the effect of out-of-plane defects such as Sr dopants; and strong-scattering disorder, parameterized by $\Gamma_{N,\mathrm{unitarity}}$, to represent native defects in the CuO$_2$ planes, such as Cu vacancies.  The combined effect of Born and unitarity-limit scattering is additive in the self energy (Eq.~\ref{tmatrix}).\cite{Nunner:2005p654}  
 
 An estimate of the scattering parameters to be used in the model can be made by comparing $\Gamma_N$  with experiment, taking care to note that the experimentally accessible scattering rate (\emph{e.g.}, that observed in an ARPES measurement of inverse lifetime\cite{Valla:1999p641}) is $2 \Gamma_N$.  Keeping this in mind and assuming for now that the momentum relaxation rate is the same as the single-particle scattering rate, the  dc resistivity due to impurity scattering will be
\begin{equation}
\rho_0 = \frac{2 \Gamma_N}{\epsilon_0 \omega_p^2} = \mu_0 \lambda_{00}^2 \times 2 \Gamma_N\,.
\label{eq:resistivity1}
\end{equation}
Here $\lambda_{00}$ is the zero-temperature penetration depth of a notional system with the same Fermi surface (doping level) that does not contain disorder.  It cannot be accessed experimentally but an estimate of $\lambda_{00}$ can be made starting from the measured zero-temperature penetration depth, $\lambda_0$, and then correcting for the degree of superfluid suppression using the $T \to 0$ limit of Eq.~\ref{superfluiddensity}:
\begin{equation}
\lambda_{00}^2 = \lambda_0^2 \Big/\frac{\rho_s(T\to 0)}{\rho_{s00}}\,.
\end{equation}
The final form for the residual resistivity is then
\begin{equation}
\rho_0  = 2 \mu_0 \lambda_0^2 \Gamma_N\Big/\frac{\rho_s(T\to 0)}{\rho_{s00}}\,.
\label{eq:resistivity2}
\end{equation}
We note that this neglects the effects of small-angle scattering, making it an \emph{upper bound} on resistivity.  If known, the small-angle scattering correction can be applied to Eq.~\ref{eq:resistivity2} as a refinement.

\section{Comparison with experiment}

\subsection{Isotropic systems}

Sufficiently far from half-filling the Fermi surface of a quasi-2D metal is well approximated by a circle, and the $d$-wave form factor is $\Omega(\phi) \approx \sqrt{2}\cos(2 \phi)$.  In this limit the angle integrals can be evaluated analytically\cite{Sun:1995jr,Huttema:2009p2230} and the Matsubara sums computed rapidly.  Results for the superfluid density are shown in Fig.~\ref{fig:CircularFS}.  The clean-limit curve displays one of the clear hallmarks of $d$-wave gap nodes: linear behaviour in $\rho_s(T)$.\cite{HARDY:1993p632}  Note that this behaviour  emerges only in the asymptotic low-temperature limit --- the substantial downwards curvature in $\rho_s(T)$ at higher temperatures is a band-structure effect due, in this case, to the particular choice of a circular Fermi surface.  

\begin{figure*}[t]
\centering
\includegraphics[width = \textwidth]{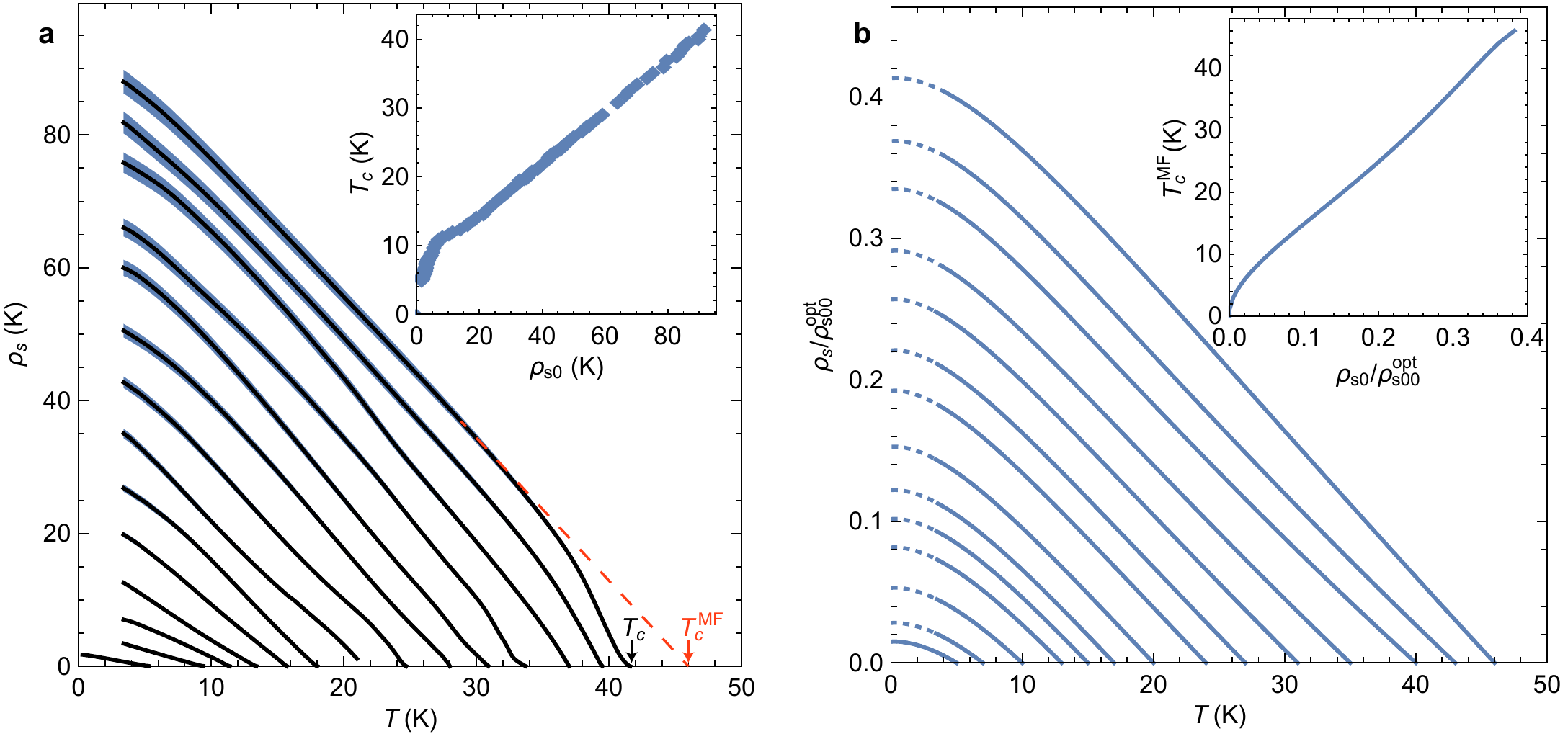}
\caption{(Color online) Superfluid density in overdoped \lsco.  (a) Main panel: Superfluid density for overdoped \lsco\ from Ref.~\onlinecite{Bozovic:2016ei}, replotted  in two dimensions, over the full temperature range that the data were measured: \mbox{$T> 300$~mK} for the lowest $T_c$ sample, $T>3.4$~K for higher $T_c$ samples. Shading indicates 2\% error bands in superfluid density (1\% error bands in penetration depth) as stated in Ref.~\onlinecite{Bozovic:2016ei}.  The dashed line shows an example of the construction used to estimate the transition temperature, $T_c^\mathrm{MF}$, to be assumed in the corresponding mean-field theory. Inset: correlation between measured superconducting transition temperature, $T_c$, and zero-temperature superfluid density, $\rho_{s0}$, replotted from Ref.~\onlinecite{Bozovic:2016ei}. (b)~Superfluid density calculated within dirty $d$-wave BCS theory, on ARPES-derived Fermi surfaces,\cite{Yoshida:2006hw} for predominantly weak scattering ($\Gamma_{N,\mathrm{Born}} = 17$~K) with a small amount of strong-scattering disorder \mbox{($\Gamma_{N,\mathrm{unitarity}} = 1$~K)}.  Main panel: Superfluid density $\rho_s$ normalized to the zero-temperature, clean-limit superfluid density for optimally doped material, $\rho_{s00}^\mathrm{opt}$.  Solid lines correspond to the temperature range accessed in the experiments, dashed lines extend this to lower temperatures. Inset: Correlation between mean-field transition temperature, $T_c^\mathrm{MF}$, and $\rho_{s0}/\rho_{s00}^\mathrm{opt}$.} 
\label{fig:Superfluid}
\end{figure*} 

As discussed above, it is convenient to parameterize the disorder level in terms of normal-state scattering rate $\Gamma_N$.  We see in Fig.~\ref{fig:CircularFS} that while $T_c$ depends only on $\Gamma_N/T_{c0}$, the form of  $\rho_s(T)$ at lower temperatures is strongly affected by the impurity phase shift.  In both the Born ($c \gg 1$) and unitarity ($c = 0$) limits there is substantial suppression of the zero-temperature superfluid density but it is only in the unitarity limit that disorder rapidly causes a cross-over to quadratic behaviour in $\rho_s(T)$ at low temperatures.\cite{Kogan:2013kf}  In contrast, it takes a large amount of Born scattering (and subsequent loss of superfluid density) before the low temperature linear behaviour in $\rho_s(T)$ is removed.  Figure~\ref{fig:CircularFS} therefore serves to illustrate that while the observation of $T^2$ behaviour in $\rho_s(T)$ is a concrete indication that disorder is important, the observation of a linear temperature dependence of $\rho_s$ does not guarantee that a material is a clean $d$-wave superconductor.

\subsection{Overdoped cuprates}

The mid-range curvature of $\rho_s(T)$ seen in Fig.~\ref{fig:CircularFS} is typically not observed in overdoped cuprate superconductors.\cite{Broun:1997p387,Panagopoulos:1999p705,Lemberger2010,Deepwell:2013uu}  To carry out a more detailed comparison with the experiments on \lsco\ requires  realistic Fermi surfaces.  The calculations presented below are based on next-next-nearest neighbour tight-binding parameterizations of $\epsilon_\mathbf{k}$ in \lsco, 
\begin{multline}
\epsilon_\mathbf{k} = \epsilon_0 - 2 t\left(\cos k_x a + \cos k_y a \right) - 4 t^\prime \cos k_x a \cos k_y a \\
- 2 t^{\prime\prime}\left(\cos 2 k_x a + \cos 2 k_y a \right)\,,
\end{multline}
obtained from fits to ARPES spectra as a function of hole doping.\cite{Yoshida:2006hw}  In the ARPES study $t = 0.25$~eV and $t^{\prime\prime}/t^\prime = -0.5$.  $t^\prime$ and $\epsilon_0$ are parameters that vary with doping, leading to the energy dispersions and Fermi surfaces shown in Fig.~\ref{fig:FermiSurfaces}.  In this model the Fermi surface is defined by $\epsilon_\mathbf{k} = 0$.  

To bridge between superfluid density and ARPES measurements we assume the standard parabolic relationship between $T_c$ and hole doping $p$.  The specific form used in Ref.~\onlinecite{Bozovic:2016ei} is 
\begin{equation}
p = 0.16+(0.01 - 2.4 \times 10^{-4}\,T_c)^{1/2}\;.
\end{equation}
This maps fairly closely onto the stated Sr concentrations in the ARPES experiment of Ref.~\onlinecite{Yoshida:2006hw}, with a slight offset: \emph{i.e.}, $x = 0.15 \to p = 0.16$ and $x = 0.22 \to p = 0.23$.  In future, it would be highly informative if ARPES measurements could be carried out on samples similar to those used in the penetration depth measurements, so that details in the electronic structure could be lined up precisely with the doping-dependent superfluid density.  In any case, we note that the parameter $p$ is a nominal hole doping, and is only used in the calculations as an internal variable.  The doping dependence of the calculated superfluid density shown in Fig.~\ref{fig:Superfluid} is not sensitive to the detailed mapping onto the ARPES experiment.  In particular, the change in Fermi surface topology as the Fermi energy passes through the van Hove point does not appear as a sharp feature in $\rho_s(p)$; this is due to the factor of $v_{F,x}^2$ in the relevant Fermi surface integral, which underweights the parts of the Fermi surface where the dispersion is flat.  The only sign of the van Hove crossing is a small cusp in the zero-temperature gap ratio, $2\Delta_0/k_B T_c$, as can be seen in the inset of Fig.~\ref{fig:Gaps}.  In the current context, the most important consequence of basing the calculations on realistic energy dispersions is the removal of spurious midrange curvature in $\rho_s(T)$.\cite{Arberg:1993cu,Xiang:1996p319}

\subsection{Disorder level}
\label{disorder}

\begin{figure}[t]
\centering
\includegraphics[width = 0.8 \columnwidth]{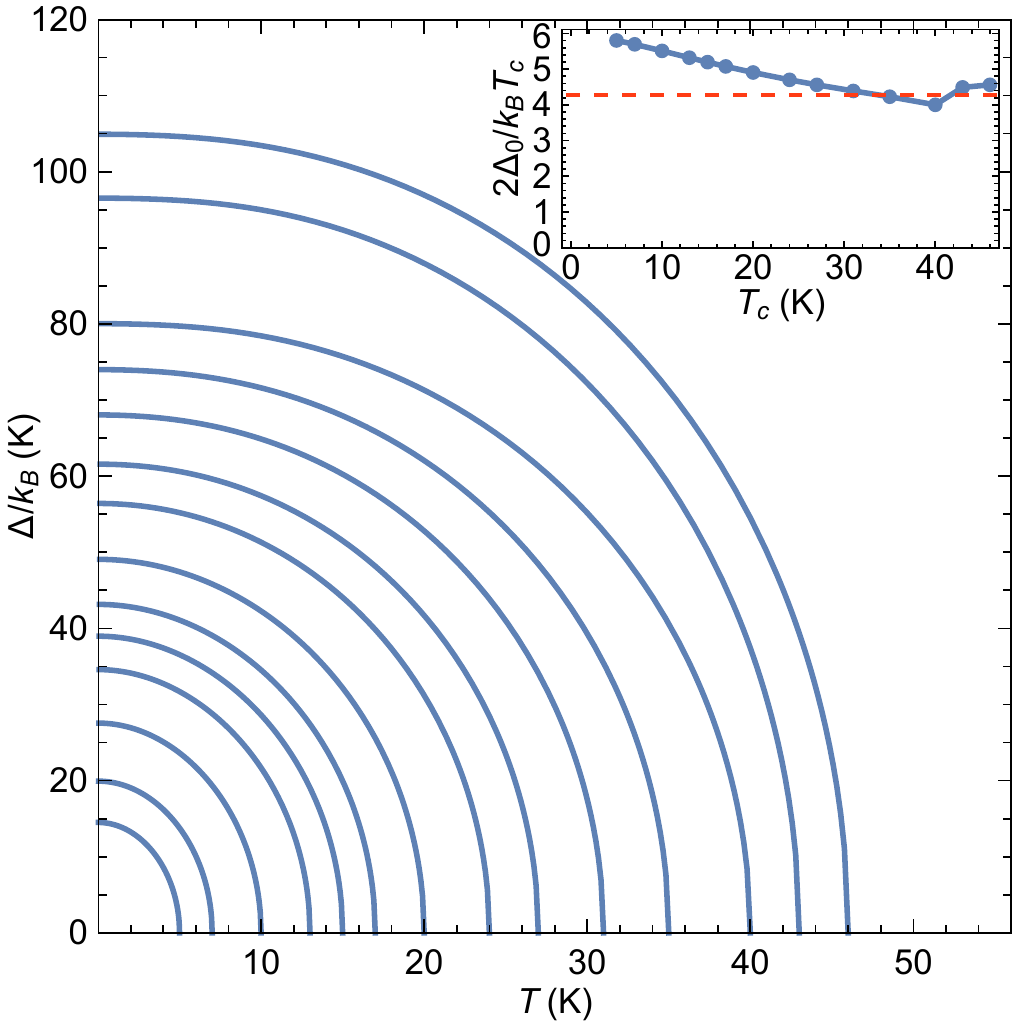}
\caption{(Color online) Superconducting gap parameter underlying the superfluid density calculation.  Main panel: Temperature and doping dependence of the gap maximum on the Fermi surface.  Inset: Doping dependence of the zero-temperature gap ratio $2 \Delta_0/k_B T_c$.  Dashed line denotes the clean-limit $d$-wave BCS value, $2 \Delta_0 = 4.28 k_B T_c$.}
\label{fig:Gaps}
\end{figure} 

\begin{figure}[t]
\centering
\includegraphics[width = 0.8 \columnwidth]{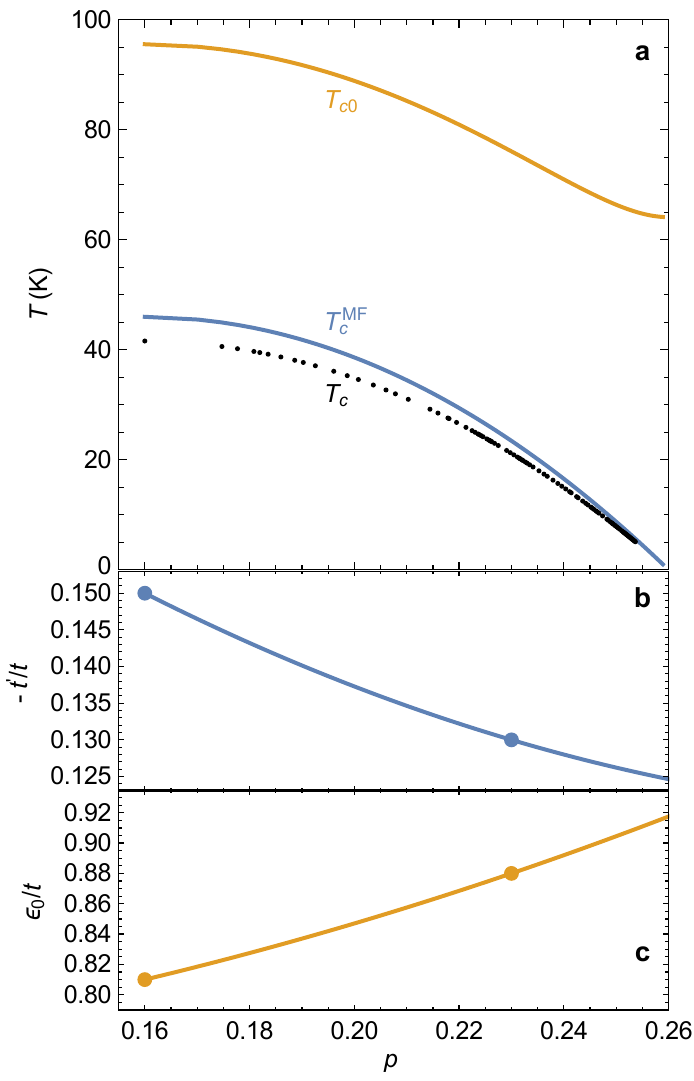}
\caption{(Color online) Doping-dependence of model parameters. (a) Clean-limit transition temperature, $T_{c0}$ and mean-field transition temperature, $T_c^\mathrm{MF}$, plotted along with the experimentally observed transition temperature, $T_c$, from Ref.~\onlinecite{Bozovic:2016ei}. (b) Next-nearest neighbour hopping integral, $t^\prime$, in units of the nearest neighbour hopping, $t$. (c) Energy offset, $\epsilon_0$, in units of $t$.  Doping-dependent tight-binding parameters are interpolations through discrete values (solid points) from Ref.~\onlinecite{Yoshida:2006hw}.}
\label{fig:Parameters}
\end{figure}

In deciding on the appropriate level of disorder to assume in the calculations, useful guidance comes from dc resistivity.  As discussed in  Sec.~\ref{theory}, the scattering rate relevant to pair-breaking and superfluid density is the elastic scattering rate, which  is related to the residual resistivity $\rho_0 \equiv \rho(T \to 0)$, not the resistivity at $T_c$.  To avoid uncertainties associated with extrapolating resistivity below $T_c$, we assume a single, doping-independent scattering rate in our calculation, as our primary aim is to illustrate the physics contained within dirty $d$-wave superconductivity.  In addition, as mentioned previously, the theoretical resistivity given in Eqs.~\ref{eq:resistivity1} and \ref{eq:resistivity2} does not account for the effects of small-angle scattering, which in the cuprates can cause the momentum relaxation rate, $\Gamma_\mathrm{tr}$, to be substantially smaller than the single-particle relaxation rate,  $\Gamma_\mathrm{sp}$.  Overdoped \tbcod\ provides a useful point of reference here, as it is sufficiently clean that quantum oscillation experiments have been performed.\cite{Vignolle:2008p1694,Bangura:2010p1675,Rourke:2010bl}  This enables the single-particle relaxation rate\cite{Bangura:2010p1675} and momentum relaxation rate\cite{Proust:P2lqZi4f} to be determined separately, with the result that $\Gamma_\mathrm{sp}/\Gamma_\mathrm{tr} = 1.7$.\cite{Deepwell:2013uu} In the absence of quantum oscillation data, we proceed by assuming the same ratio for \lsco.  This is not unreasonable, as the dominant source of disorder in both systems is out-of-plane cation disorder: in the case of \tbcod, an excess of Cu atoms, occupying Tl sites;\cite{Shimakawa:1993jm} and, in \lsco, the partial substitution of Sr for La that is an inherent part of its hole-doping mechanism.  With these assumptions, and taking a residual resistivity $\rho_0 \approx 16~\mu\Omega$\,cm, we obtain a normal-state scattering rate $\Gamma_N \approx 18$~K.  In the calculations presented here we partition this between predominantly weak, Born-limit scattering ($\Gamma_{N,\mathrm{Born}} = 17$~K) with a small amount of strong, unitarity-limit scattering ($\Gamma_{N,\mathrm{unitarity}} = 1$~K).  The inclusion of a small amount of strong-scattering disorder causes a low-temperature crossover to $T^2$ dependence in $\rho_s(T)$, which is hinted at by the data from Ref.~\onlinecite{Bozovic:2016ei}, plotted in Fig.~\ref{fig:Superfluid}(a).  Note that this choice of scattering rate satisfies the conditions for clean-limit superconductivity, $\Gamma_\mathrm{tr} \ll 2 \Delta_0$, over most of the doping range.  In fact, based on the scattering rate estimates above and the gap values plotted in Fig.~\ref{fig:Gaps}, we estimate $\Gamma_\mathrm{tr} \sim 2 \Delta_0$ only for $T_c < 5$~K.  Nevertheless, there is substantial loss of superfluid density to disorder across the entire doping range, even where the conventional criteria firmly place superconductivity in the clean limit.  This points to pair breaking in $d$-wave superconductors being a process that spreads uncondensed spectral weight over a wide range of sub-gap frequencies.

\section{Discussion}

The calculated superfluid density is presented in Fig.~\ref{fig:Superfluid}(b).  In our calculations the underlying Fermi surface follows the doping dependence of the electronic structure measured in the ARPES experiments\cite{Yoshida:2006hw} \emph{and is not an adjustable parameter.}  At each value of the nominal hole doping $p$, the underlying clean-limit transition temperature, $T_{c0}$, is set according to Eq.~\ref{Tc_equation} so that $T_c$, the transition temperature in the presence of disorder, matches the mean-field transition temperature, $T_c^\mathrm{MF}$, inferred by extrapolating the linear regime of the experimental $\rho_s(T)$ curve to zero, as shown in Fig.~\ref{fig:Superfluid}(a).   The doping dependences of $T_c$,  $T_c^\mathrm{MF}$ and $T_{c0}$ are plotted in Fig.~\ref{fig:Parameters}(a) and the tight-binding parameters used in the calculation are shown in Figs.~\ref{fig:Parameters}(b) and \ref{fig:Parameters}(c).  As described in the previous section, the amount of impurity scattering is the only independent parameter in the theory and has been fixed as a function of doping, for the sake of simplicity, as discussed in Sec.~\ref{disorder}.
 
The superfluid density  calculated from dirty $d$-wave BCS theory reproduces many of the features observed in the experiments in Ref.~\onlinecite{Bozovic:2016ei}. In particular, $\rho_s(T)$ shows a strong, nearly linear temperature dependence over almost the full doping range, despite the strong suppression of superfluid density caused by the disorder.  The correlation between $T_c$ and $\rho_{s0}$ [see inset of Fig.~\ref{fig:Superfluid}(b)] also reproduces the main features of the experiments, namely the almost linear dependence at higher $T_c$, with finite intercept, crossing over to square-root behaviour at low $T_c$.  Indeed, very similar behaviour of $T_c(\rho_{s0})$ was found in earlier calculations of superfluid density for Born-limit scattering on a \emph{circular} Fermi surface (Fig.~4 of Ref.~\onlinecite{Puchkaryov1998}).  

One concern might be that the crossover in $T_c(\rho_{s0})$ from linear to square-root behaviour does not occur as smoothly in the experimental data as in the theoretical curve.  In the idealized form considered in this paper, the dirty $d$-wave BCS theory assumes spatial homogeneity, whereas there is a body of evidence that real samples of \lsco\ consist of an inhomogeneous mixture of superconducting and metallic regions, with the fraction of metallic regions increasing on overdoping.\cite{Tanabe:2005km,Wang2007}  While the samples in Ref.~\onlinecite{Bozovic:2016ei} are grown with exquisitely controlled \emph{average} composition, the doping mechanism in \lsco\ is based on random substitution of cations and inhomogeneity must always become relevant when approaching the overdoped phase boundary.  In addition, the technique used to characterize inhomogeneity in Ref.~\onlinecite{Bozovic:2016ei} (measurement of the temperature width of the out-of-phase susceptibility near $T_c$) is not a good probe of in-plane microscopic inhomogeneity, as the superconducting coherence length diverges at $T_c$, averaging over short-length-scale inhomogeneity.  For an inhomogeneous superconductor the electrodynamic response can be modelled using effective medium theory of the conductivity.\cite{Cohen:1973jg,Waldram:1999p2205} This is described in Appendix~\ref{app:effectiveMedium}, where the macroscopic superfluid density is shown to simply be a scaled version of the microscopic superfluid density in the superconducting regions.  The effect of this type of inhomogeneity would be to distort the theoretical $T_c(\rho_{s0})$ curve to the left as the inhomogeneous regime is entered, which may explain the kink at low $\rho_{s0}$ that appears in the experimental curve [see inset of Fig.~\ref{fig:Superfluid}(a)].

Another possible concern is the degree of superfluid suppression predicted by the calculations, which at first sight appears surprising large.  We emphasize again that the disorder level assumed in the calculation corresponds closely to the observed resistivity, and is really the only adjustable parameter in the model.  Here, a useful cross-check comes again from comparison with overdoped  \tbcod, where for $T_c \approx 25$~K material the degree of superfluid suppression, $\rho_{s0}/\rho_{s00}$, is estimated to lie in the range 0.25 to 0.4,\cite{Deepwell:2013uu} despite overdoped \tbcod\ having a residual resistivity\cite{Proust:P2lqZi4f} less than half that of \lsco.

Finally, while the main purpose of our calculation is to illustrate the qualitative features contained within dirty $d$-wave superconductivity,  it is interesting that the implied clean-limit transition temperatures, $T_{c0}$, are large and have a suggestive similarity to the temperatures at which the first experimental signatures of superconductivity are observed in properties such as magnetoresistance.\cite{Rourke2011} The mean-field model considered here does not include fluctuation effects, which are known to be important in the cuprates\cite{Corson:1999p716,Xu:2000p609,Wang:2002p645,Wang:2005p2400,Wang:2006p185,Kitano:2006id,Li:2007p180,Grbic:2009kd,Rourke2011,Bilbro:2011jj,Grbic:2011hp,Behnia:2016fc} and are probably responsible for the downturns in the experimentally observed $\rho_s(T)$ on the approach to $T_c$.\cite{KAMAL:1994p701}  Nevertheless, one way in which the underlying $T_{c0}$ might become visible in experiments would be as rare regions in which the local disorder level is lower than average.  The overall implication is that disorder not only plays a role in limiting superfluid density in the overdoped cuprates, but in suppressing the transition temperature.  This suggests that it may be possible to enhance  $T_c$, as well as the doping range over which superconductivity occurs, by controlled engineering of disorder in these materials.

\section{Conclusions}

In conclusion, we find that dirty $d$-wave BCS theory, applied to a realistic parameterization of the doping-dependent Fermi surface,  reproduces most of the phenomenology of the superfluid density in overdoped \lsco.\cite{Bozovic:2016ei}   A strong suppression of  superfluid density is achieved without introducing significant curvature in $\rho_s(T)$ by considering predominantly weak, Born-limit scattering, at a disorder level compatible with the observed resistivity, and in a regime that firmly satisfies the conventional definition of  clean-limit superconductivity, $\Gamma_\mathrm{tr} \ll 2 \Delta_0$.   We conclude that the correlation between $T_c$ and $\rho_s$ observed in the overdoped cuprates is a generic feature of a disordered $d$-wave superconductor.

\acknowledgements

We acknowledge useful discussions with N.~Doiron-Leyraud, P.~J.~Hirschfeld, F.~Lalibert\'e and L.~Taillefer.   We gratefully acknowledge financial support from the Natural Science and Engineering Research Council of Canada, the Canadian Institute for Advanced Research, and the Canadian Foundation for Innovation.

\clearpage

\appendix

\section{Effective-medium theory of an inhomogeneous superconductor}
\label{app:effectiveMedium}

One way to account for the effects of microscopic inhomogeneity is effective medium theory.\cite{Cohen:1973jg}  This has been carried out in Ref.~\onlinecite{Waldram:1999p2205} for a two-dimensional system consisting of normal regions of conductivity $\sigma_n$ and superconducting regions of conductivity $\sigma_s$.  In the case of the low-frequency superfluid density it is useful to first consider the limit of nonzero frequency, $\omega$, where the conductivity of the superconductor is finite and predominanly imaginary, $\sigma_s \approx 1/\I \omega \mu_0 \lambda^2$.  According to the theory, the effective conductivity, $\sigma$, is a root of the equation
\begin{equation}
\sigma^2 + (1 - 2f)(\sigma_s - \sigma_n) \sigma - \sigma_n \sigma_s = 0\,,
\end{equation}
where $f$ is the fraction of the sample occupied by the superconducting regions.  In the low frequency limit $\sigma_s \gg \sigma_n$. Then
 \begin{equation}
\sigma \to (2 f -1)\sigma_s = (2 f - 1)\frac{1}{\I \omega \mu_0 \lambda^2}\;.
\end{equation}
The static superfluid density is defined to be 
\begin{equation}
\rho_s \equiv \lim_{\omega \to 0} \omega \mu_0 \mbox{Im}\left\{\sigma\right\} = (2 f - 1) \frac{1}{\lambda^2}\,.
\end{equation}
We therefore expect the macroscopic (observed) superfluid density to be a scaled version of the microscopic superfluid density, with the scale factor ranging from 1 in the limit of no inhomogeneity to 0 on approach to the percolation threshold at $f = \frac{1}{2}$.


%

\newpage

\onecolumngrid

\begin{center}
{\bf Erratum: Disorder and  superfluid density in overdoped cuprate superconductors}
\end{center}

In Section III~C of our paper, the scattering parameters were reported as $\rho_0 \approx 16~\mu\Omega$cm, $\Gamma_N = 18$, $\Gamma_{N,\mathrm{Born}} = 17$ and $\Gamma_{N,\mathrm{unitarity}} = 1$~K.  The actual values used in the calculations and the preparation of the figures were a factor of $\pi$ larger; i.e., $\rho_0 \approx 50~\mu\Omega$cm, $\Gamma_N = 18 \pi$, $\Gamma_{N,\mathrm{Born}} = 17 \pi$ and $\Gamma_{N,\mathrm{unitarity}} = \pi$~K.  Note that the revised value of $\rho_0$ is in good correspondence with the residual terahertz conductivities reported in Fig.~S13 of Ref.~\onlinecite{mahmood:2017}. In Fig.~4, the plot of gap magnitude, $\Delta$, showed the prefactor of the $d$-wave form factor $\cos(k_x a) - \cos(k_y a)$ rather than the gap maximum on the Fermi surface, which is replotted below.

\setcounter{figure}{3}

\begin{figure*}[h]
\centering
\includegraphics[width = 0.5 \columnwidth]{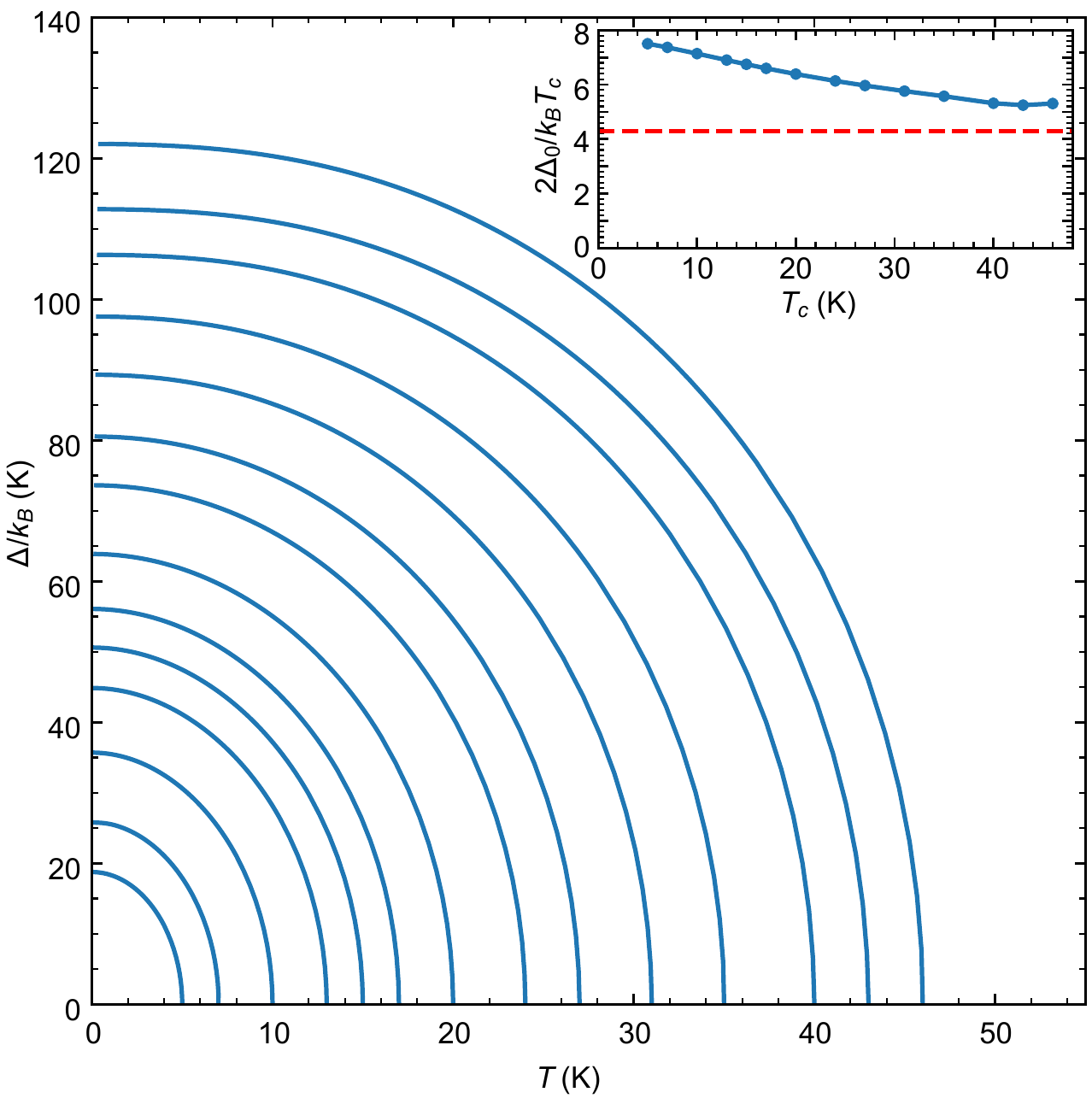}
\caption{(Color online) Revised plot of the gap magnitude, $\Delta$, which now correctly shows the $d$-wave gap maximum on the \lsco\ tight-binding Fermi surface.  Inset: Revised doping dependence of the zero-temperature gap ratio $2 \Delta_0/k_B T_c$, where $\Delta_0$ is now the Fermi-surface gap maximum at zero temperature.  Dashed line denotes the clean-limit $d$-wave BCS value for a circular Fermi surface, $2 \Delta_0 = 4.28 k_B T_c$.}
\label{fig:Gaps}
\end{figure*} 

We note that these corrections do not affect the other figures, results, and conclusions of our paper.


%

\end{document}